\documentclass[preprint,12pt]{elsarticle}

\usepackage{amssymb}
\usepackage{color}

\newcommand{\alkor}[1]{\textcolor{blue}{#1}}
\renewcommand{\alkor}[1]{#1}

\journal{Chaos, Solitons and Fractals}

\begin{document}

\begin{frontmatter}

\title{Generalized synchronization in discrete maps. \\ New point of view on weak and strong synchronization}


\author[SSU]{Alexey~A.~Koronovskii}
\author[SSU]{Olga~I.~Moskalenko\corref{lulu}}
\cortext[lulu]{Corresponding Author} \ead{moskalenko@nonlin.sgu.ru}
\author[SSU]{Svetlana~A.~Shurygina}
\author[SSU]{Alexander~E.~Hramov}
\address[SSU]{Faculty of Nonlinear Processes, Saratov State University, 83, Astrakhanskaya, Saratov,
410012, Russia}

\begin{abstract}
In the present Letter we show that the concept of the generalized
synchronization regime in discrete maps needs refining in the same
way as it has been done for the flow systems [PRE, \textbf{84}
(2011) 037201]. We have shown that\alkor{, in the general case, when the relationship between state vectors of the
interacting chaotic maps are considered,} the
prehistory must be taken into account. We extend the phase tube
approach to the systems with a discrete time coupled both
unidirectionally and mutually and analyze the essence of the
generalized synchronization by means of this technique. Obtained
results show that the division of the generalized synchronization
into the weak and the strong ones also must be reconsidered.
Unidirectionally coupled logistic maps and H\'enon maps coupled
mutually are used as sample systems.
\end{abstract}


\end{frontmatter}

\section{Introduction}
\label{sct:Introduction}

Chaotic synchronization of nonlinear dynamical systems is an
universal phenomenon having a large fundamental significance and
different practical applications in all fields of science and
technique~\cite{Glass:2001_SynchroBio, Boccaletti:2002_ChaosSynchro,
alkor:2010_SecureCommunicationUFNeng}. The presence of synchronous
behavior can be observed in different mathematical, physical,
sociological, physiological, biological and other systems. There are
a lot of types of chaotic synchronization such as complete, phase,
generalized, noise-induced, lag and time scale synchronization.

One of the most interesting types of the synchronous chaotic system
behavior is the generalized
synchronization~\cite{Rulkov:1995_GeneralSynchro}. This type of
chaotic synchronization is traditionally introduced for two
unidirectionally coupled flow chaotic
oscillators~\cite{Rulkov:1995_GeneralSynchro,Harmov:2005_GSOnset_EPL},
spatially distributed media~\cite{Hramov:2005_GLEsPRE,
Filatov:2006_PierceDiode_PLA, dmitriev:074101} or discrete
maps~\cite{Pyragas:1996_WeakAndStrongSynchro,Aeh:2005_GS:ModifiedSystem}
and means the presence of the \emph{functional relation} between the
drive and response system states. This functional relation is
supposed to be smooth or
fractal~\cite{Pyragas:1996_WeakAndStrongSynchro}, although there are
no technique to find the implicit form of this relation (except for
the complete and lag synchronization regimes). In the framework of
the existing concept accepted generally, the strong and weak types
of the generalized synchronization may be distinguished, according
to the properties of the functional relation. Strong synchronization
is assumed to correspond to the smooth map between variables of the
drive and response systems (this regime is supposed to be observed
in the case of complete and lag synchronization), whereas the weak
one means the existence of a fractal map between them and can be
detected with the help of an auxiliary system
approach~\cite{Rulkov:1996_AuxiliarySystem}.

Recently, we have refined the concept of generalized synchronization
in the flow systems and shown that the state vectors of the
interacting chaotic systems \alkor{should be considered as} related with each other by the
\emph{functional} \alkor{instead of} the \emph{functional relation}~\cite{Koronovskii:GSTubePRE2011}. We have also
proposed the \emph{phase tube} approach explaining the essence of
generalized synchronization and allowing the detection of the
generalized synchronization regime in many relevant physical
circumstances including bidirectionally coupled chaotic
oscillators~\cite{Koronovskii:GSTubePRE2011}.

Now, we have to make the next important step. The notion of
generalized synchronization has been introduced for chaotic
oscillators irrelatively of the type of the oscillators, it covers
both the flow systems and discrete
maps~\cite{Pyragas:1996_WeakAndStrongSynchro,Aeh:2005_GS:ModifiedSystem,
Hramov:2006_PLA_NIS_GS}. At the same time, the approach proposed
in~\cite{Koronovskii:GSTubePRE2011} has been developed only for the
flow systems. In the present Letter we extend the phase tube
approach on discrete maps coupled both unidirectionally and
mutually. As it would be shown bellow, \alkor{in the general case}
the relation between states of the interacting discrete maps being
in the generalized synchronization regime is analogous to the
\emph{functional} (as it takes place in the flow systems). As a
consequence, the concepts of the weak and strong synchronization of
chaos must also be reconsidered.

\section{The theory of generalized synchronization for discrete maps}
\label{sct:GS}

First of all, based on the results of our previous
work~\cite{Koronovskii:GSTubePRE2011} for the flow systems, we
briefly describe the refined theory of the generalized
synchronization regime for the discrete maps.

The definition of the generalized synchronization regime generally
accepted hitherto is the presence of a functional relation
\begin{equation}
{\mathbf{y}=\mathbf{F}[\mathbf{x}]} \label{eq:FunctRelGeneral}
\end{equation}
between the drive $\mathbf{x}$ and response $\mathbf{y}$ oscillator
states~\cite{Rulkov:1995_GeneralSynchro,
Pyragas:1996_WeakAndStrongSynchro}. Obviously,
Eq.~(\ref{eq:FunctRelGeneral}), as  applied to maps, should be
written in the form
\begin{equation}
{\mathbf{y}_{n}=\mathbf{F}[\mathbf{x}_{n}]} \label{eq:FunctRelMaps},
\end{equation}
where $\mathbf{x_n}$ and $\mathbf{y_n}$ are the drive and response
maps, respectively. The evolution of the vectors $\mathbf{x_n}$ and
$\mathbf{y_n}$ is determined by
\begin{equation}\label{eq:EvolutionOperators}
\begin{array}{l}
\mathbf{x}_{n+1}=\mathbf{H}(\mathbf{x}_{n},\mathbf{g}_x),\\
\mathbf{y}_{n+1}=\mathbf{G}(\mathbf{y}_{n},\mathbf{g}_y)+
\sigma\mathbf{P}(\mathbf{x}_{n},\mathbf{y}_{n}),\\
\end{array}
\end{equation}
where $\mathbf{H}$ and $\mathbf{G}$ are the evolution operators of
the considered discrete systems, $\mathbf{g}_x$ and $\mathbf{g}_y$
are the controlling parameter vectors, $\mathbf{P}$ denotes the
coupling term and $\sigma$ is the scalar coupling parameter. Without
the lack of generality we shall suppose below the identical
dimension $m$ of the phase space of the drive and response systems.

In our work~\cite{Koronovskii:GSTubePRE2011} we have shown for the
flow systems that $\mathbf{F}[\cdot]$ in
Eq.~(\ref{eq:FunctRelGeneral}) \alkor{should} be considered as
\emph{a functional} (\alkor{contrary to} \emph{a functional
relation}), that means that the system state $\mathbf{y}(t)$ depends
not only on the state of the drive system $\mathbf{x}(t)$ but on the
prehistory with the length $\tau$ of the drive oscillator
$\mathbf{x}(s)$, ${t-\tau<s\leq t}$. \alkor{From the theoretical
point of  view, for two coupled \emph{flow} systems the existence of
the functional relation~(\ref{eq:FunctRelGeneral}) can be proven
rigorously~\cite{Kocarev:1996_GS} only for the \emph{unidirectional}
type of coupling, and, in the most cases, this functional relation
is not continuously differentiable. As far as the mutual coupled
flow systems are concerned, the theoretical proof mentioned above
becomes inapplicable. The consideration of the generalized
synchronization of flow systems from the point of view of \emph{a
functional} allows to avoid both the uncertainty of the functional
relation existence and the nondifferentiability feature.}

\alkor{Since the flow systems may be reduced to the discrete maps
with the help of the of Poincar\'{e} secant approach (see, e.g.,
\cite{Filatova:MapsFlowsJETP2005}), the functional relation
existence between system states for the generalized synchronization
regime may be extended only for \emph{unidirectionally coupled
invertible} maps, and, again, this functional relation is
\emph{fractal} (i.e., it is not continuously differentiable)
typically. As far as the the non-invertible and mutually coupled
maps are concerned, the existence of the functional relation does
not seem to be rigorously true. Therefore, having considered all
mentioned above, one can come to conclusion that the prehistory
should be considered in the same way, as it had been done for the
flow systems~\cite{Koronovskii:GSTubePRE2011}.} In terms of the
discrete maps this circumstance may be taken into account by the
following modification of Eq.~(\ref{eq:FunctRelMaps})
\begin{equation}
{\mathbf{y}_{n}=\mathbf{F}[\mathbf{x}_{n}, \mathbf{x}_{n-1}, \dots, \mathbf{x}_{n-K}]} \label{eq:FunctionalMaps},
\end{equation}
where $K$ is the discrete length of the prehistory being sufficient
for the unique determination of the state of the response map
$\mathbf{y}_n$.

Let $\mathbf{x}_N$ and $\mathbf{y}_N$ be the reference points
belonging to the chaotic attractors of the drive and response maps
being in the generalized synchronization regime, respectively. Let
also $\delta \mathbf{y}_{Jk}=\mathbf{y}_{J-k}-\mathbf{y}_{N-k}$ and
$\delta \mathbf{x}_{Jk}=\mathbf{x}_{J-k}-\mathbf{x}_{N-k}$
(${k=0,\dots,K}$), be the vectors characterizing the deviation of
the trajectories under consideration $\mathbf{x}_{J-k}$,
$\mathbf{y}_{J-k}$ from the reference trajectories
$\mathbf{x}_{N-k}$ and $\mathbf{y}_{N-k}$. For the neighbor point
$\mathbf{x}_J$ of the drive oscillator such that
${||\delta\mathbf{x}_{J}||=||\delta\mathbf{x}_{J0}||<\varepsilon}$
its image $\mathbf{y}_J$ in the response system is also close to the
reference point $\mathbf{y}_N$ (see
\cite{Rulkov:1995_GeneralSynchro} for detail), i.e.,
${||\delta\mathbf{y}_{J}||=||\delta\mathbf{y}_{J0}||<\delta(\varepsilon)}$.
Having supposed that
\begin{equation}\label{eq:InfinitesimalityAssumption}
{||\delta\mathbf{x}_{Jk}||<\varepsilon}, \quad k=0,\dots, K
\end{equation}
and linearized Eq.~(\ref{eq:FunctionalMaps}), one obtains that
\begin{equation}\label{eq:Linearization}
\delta\mathbf{y}_{J}=\sum\limits_{k=0}^K J_{\mathbf{x}_{N-k}}\mathbf{F}[\mathbf{x}_{N},\dots,\mathbf{x}_{N-K}]\delta \mathbf{x}_{Jk},
\end{equation}
where $J_{\mathbf{x}_{N-k}}$ is the Jacobian operator for $k$-th
variable. Since the form of $\mathbf{F}[\cdot]$ can not be found
explicitly, Eq.~(\ref{eq:Linearization}) may be rewritten in the
form
\begin{equation}\label{eq:MatrixForm}
\delta\mathbf{y}_J=\sum\limits_{k=0}^K \mathbf{A}_k\delta\mathbf{x}_{Jk},
\end{equation}
where
$\mathbf{A}_k=J_{\mathbf{x}_{N-k}}\mathbf{F}[\mathbf{x}_{N},\dots,\mathbf{x}_{N-K}]$
($k=0,\dots,K$)  are the unknown matrixes. Obviously, the
coefficients of $\mathbf{A}_k$-matrix are determined by the whole
set of the vectors $\mathbf{x}_{N-K},\dots,\mathbf{x}_{N}$, but,
since the elements of this sequence are uniquely connected with each
other by the evolution operator~(\ref{eq:EvolutionOperators}), one
can assume that $\mathbf{A}_k$ depends only on $\mathbf{x}_{N-K}$,
i.e., $\mathbf{A}_k=\mathbf{A}_k(\mathbf{x}_{N-K})$.

Under assumption~(\ref{eq:InfinitesimalityAssumption}) made above,
in view of the linearity, one can write
\begin{equation}
\delta\mathbf{x}_{Jk}=\mathbf{B}_k(\mathbf{x}_{N-K})\delta\mathbf{x}_{J}
\end{equation}
[where $\mathbf{B}_k(\mathbf{x}_{N-K})$ is the unknown
matrix\footnote{Except for
$\mathbf{B}_0(\mathbf{x}_{N-K})=\mathbf{E}$, where $\mathbf{E}$ is
the identity matrix.} whose coefficients depend both on the
reference vector $\textbf{x}_{N-K}$ and the number $k$ of the
considered deviation $\delta\mathbf{x}_{Jk}$], which results in
\begin{equation}
\delta\mathbf{y}_J=\sum\limits_{k=0}^K \mathbf{A}_k(\mathbf{x}_{N-K})\mathbf{B}_k(\mathbf{x}_{N-K})\delta\mathbf{x}_{J},
\end{equation}
and, as a consequence, in
\begin{equation}\label{eq:VecorRelationship}
\delta\mathbf{y}_J=\mathbf{C}\delta\mathbf{x}_{J},
\end{equation}
where $\mathbf{C}$ is the matrix defined as
\begin{equation}
\mathbf{C}=\sum\limits_{k=0}^K \mathbf{A}_k(\mathbf{x}_{N-K})\mathbf{B}_k(\mathbf{x}_{N-K}).
\end{equation}

Note, also, within the framework of the traditional concept of the
generalized synchronization implying that the states of the
interacting systems are connected with each other by \emph{continuously differentiable} functional
relation~(\ref{eq:FunctRelMaps}) one can obtain the similar
to~(\ref{eq:VecorRelationship}) relationship
\begin{equation}\label{eq:VecorFuncRelations}
\delta\mathbf{y}_J=\mathbf{\tilde{C}}\delta\mathbf{x}_{J},
\end{equation}
with the only one difference that
\begin{equation}
\mathbf{\tilde{C}}=J\mathbf{F}[\mathbf{x}_n].
\end{equation}

Despite of the similarity of Eq.~(\ref{eq:VecorRelationship}) and
Eq.~(\ref{eq:VecorFuncRelations}) there is a great difference
between them. Indeed, Eq.~(\ref{eq:VecorRelationship}) has been
obtained under assumption that the phase trajectories
$\mathbf{x}_{N-k}$ and $\mathbf{x}_{J-k}$ (${k=0,\dots,K}$) are
close to each other on the whole prehisory time interval with the
length $K$ (see Eq.~(\ref{eq:InfinitesimalityAssumption})), whereas
Eq.~(\ref{eq:VecorFuncRelations}) requires only the nearness of two
points, $\mathbf{x}_{N}$ and $\mathbf{x}_{J}$, i.e., instead of
Eq.~(\ref{eq:InfinitesimalityAssumption}) it requires only
\begin{equation}\label{eq:OneInfinitesimalityAssumption}
{||\delta\mathbf{x}_{J}||<\varepsilon}.
\end{equation}

\begin{figure}
\centerline{\includegraphics[scale=0.75]{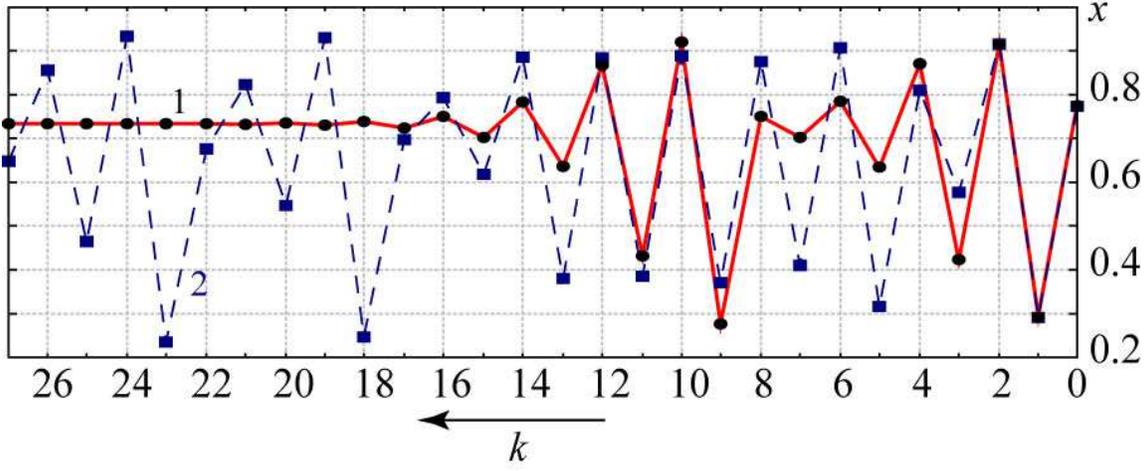}}
\caption{The dependencies of $x$-value of logistic map~(\ref{eq:LogisicMap}) on the prehisory time $k$. The reference trajectory $\mathbf{x}_{N-k}$ is shown by bold line, two trajectories $\mathbf{x}_{J1-k}$ $\mathbf{x}_{J2-k}$ (whose end points $\mathbf{x}_{J1}$ $\mathbf{x}_{J2}$ are close to $\mathbf{x}_{N}$) are shown by points $\bullet$ (line~1) and dashed line with points \textcolor{blue}{$\blacksquare$} (line~2), respectively. The horizontal axis is shown in the opposite direction}
\label{fgr:Trajectories}
\end{figure}

Since for the chaotic systems the phase trajectories can both
converge and diverge, the nearness of $\mathbf{x}_{N}$ and
$\mathbf{x}_{J}$ (Eq.~(\ref{eq:OneInfinitesimalityAssumption})) does
not mean the fulfilment of the
requirement~(\ref{eq:InfinitesimalityAssumption}), i.e., among the
vectors $\mathbf{x}_J$ being close to $\mathbf{x}_{N}$ only small
part of them satisfies the
requirement~(\ref{eq:InfinitesimalityAssumption}). This statement is
illustrated in Fig.~\ref{fgr:Trajectories} for the logisic map
\begin{equation}\label{eq:LogisicMap}
x_{n+1}=a x_n(1-x_n), \qquad a=3.75.
\end{equation}
One can see that, although both the points ${x}_{J1}$  and
${x}_{J2}$ are close to the reference state ${x}_N$ (and for both of
them requirement Eq.~(\ref{eq:OneInfinitesimalityAssumption}) is
fulfilled), only the point ${x}_{J1}$ obeys
Eq.~(\ref{eq:InfinitesimalityAssumption}) due to the nearness of the
whole trajectory ${x}_{J1-k}$ ($\bullet$) to ${x}_{N-k}$, whereas
for the point ${x}_{J2}$ (\textcolor{blue}{$\blacksquare$})
condition~(\ref{eq:InfinitesimalityAssumption}) fails, since its
trajectory ${x}_{J2-k}$ is not close to the reference one
${x}_{N-k}$ on the whole prehistory interval with the length $K$.

Although the coefficients of the matrixes $\mathbf{C}$ and
$\mathbf{\tilde{C}}$ are unknown, the validity of both
Eq.~(\ref{eq:VecorRelationship}) and
Eq.~(\ref{eq:VecorFuncRelations}) may be verified if there are $N>m$
nearest neighbors $\mathbf{x}_{Ji}$ of the reference vector
$\mathbf{x}_N$ and corresponding them vectors $\mathbf{y}_{Ji}$ of
the response map. Note also, all vectors $\mathbf{x}_{Ji}$ being
close to $\mathbf{x}_N$ can be used to check
Eq.~(\ref{eq:VecorFuncRelations}), whereas for the examination of
Eq.~(\ref{eq:VecorRelationship}) only vectors $\mathbf{x}_{Ji}$ are
applicable whose prehistory trajectory satisfies
Eq.~(\ref{eq:InfinitesimalityAssumption}).

Having tested the presence of the generalized synchronization (e.g.,
with the help of the auxiliary system approach) we can pick out $m$
nearest neighbors $\mathbf{x}_{Ji}$ (${i=1,\dots,m}$) and
corresponding to them vectors $\mathbf{y}_{Ji}$ to determine the
coefficients of the matrix $\mathbf{C}$ (or $\mathbf{\tilde{C}}$)
with the help of Eq.~(\ref{eq:VecorRelationship}) (or
Eq.~(\ref{eq:VecorFuncRelations})), respectively) in the same way as
it has been done in~\cite{Koronovskii:GSTubePRE2011}. Afterwards,
having determined the coefficients of the matrix $\mathbf{C}$ (or
$\mathbf{\tilde{C}}$) we can now find the vectors
$\delta\mathbf{z}_{Ji}$, (${i=m+1,\dots,N}$) as
\begin{equation}\label{eq:Zvectors}
\delta\mathbf{z}_{Ji}= \mathbf{C}\delta\mathbf{x}_{Ji}\quad\mbox{or}\quad\delta\mathbf{z}_{Ji}= \mathbf{\tilde{C}}\delta\mathbf{x}_{Ji}
\end{equation}
and compare them with the vectors $\delta\mathbf{y}_{Ji}$ of the
response system to check Eq.~(\ref{eq:VecorRelationship}) (or
Eq.~(\ref{eq:VecorFuncRelations}). To characterize the degree of
closeness of the vectors $\delta\mathbf{y}_{Ji}$ and
$\delta\mathbf{z}_{Ji}$ with each other one can compute the
normalized differences
\begin{equation}
\Delta_{Ji}=\frac{||\delta\mathbf{y}_{Ji}-\delta\mathbf{z}_{Ji}||}{||\delta\mathbf{y}_{Ji}||}
\label{eq:NormDif}
\end{equation}
for each pair of vectors and build their distributions.

So, the strategy of the investigation of the generalized
synchronization essence may be the following. Firstly,
Eq.~(\ref{eq:VecorFuncRelations}) must be checked for the set of
vectors $\mathbf{x}_{Ji}$ being nearest to the reference one
$\mathbf{x}_{N}$, i.e., all points satisfying
requirement~(\ref{eq:OneInfinitesimalityAssumption}) must be used.
If Eq.~(\ref{eq:VecorFuncRelations}) is valid, it means that in the
generalized synchronization regime the states of discrete maps are
connected with each other by the functional
relation~(\ref{eq:FunctRelMaps}). Alternatively, the violation of
Eq.~(\ref{eq:VecorFuncRelations}) indicates that
Eq.~(\ref{eq:FunctRelMaps}) being the main definition of the
generalized synchronization concept accepted hitherto should be reconsidered. In this case the second step consists in
the verification of Eq.~(\ref{eq:VecorRelationship}) (and
Eq.~(\ref{eq:FunctionalMaps}), respectively) with the help of the
consideration only vectors  $\mathbf{x}_{Ji}$ whose trajectories
$\mathbf{x}_{Ji-k}$ satisfy the
requirement~(\ref{eq:InfinitesimalityAssumption}), with the rest of
the vectors $\mathbf{x}_{Ji}$ used previously to check
Eq.~(\ref{eq:VecorFuncRelations}) having to be eliminated from the
consideration\footnote{For the flow system this procedure has been
named as \emph{the phase tube}
approach~\cite{Koronovskii:GSTubePRE2011}.}. Since the length $K$ of
the prehistory (or the length of the phase tube) is inversely
proportional to the absolute value of the largest conditional
Lyapunov exponent $\lambda_1^r<0$, it may be estimated as $K\sim
1/{|\lambda_1^r|}$.

In this Letter the generalized synchronization in the discrete maps
is studied for two sample systems: two unidirectionally coupled
logistic maps and two mutually coupled H\'enon maps. As we  will see
below, the concept of the generalized synchronization for the
discrete maps needs refining in the same way as it has been done for
the flow systems, since\alkor{, in the general case, for} the state vectors of the interacting chaotic
\alkor{maps the prehistory must
be taken into account}. As a consequence, the division of generalized
synchronization into weak and strong ones must also be reconsidered.
At the same time, fortunately, this modification of the generalized
synchronization concept does not discard the majority of the
obtained hitherto results concerning generalized synchronization.

\section{Logistic maps}
\label{sct:Logistic}

As the first example we consider two unidirectionally coupled
logistic maps:
\begin{equation}
\begin{array}{l}
x_{n+1}=f(x_n,a_x),\\
y_{n+1}=f(y_n,a_y)+\sigma(f(x_n,a_x)-f(y_n,a_y)),
\end{array}
\label{eq:LogMaps}
\end{equation}
where $f(x,a)=ax(1-x)$, $a_x=3.75$, $a_y=3.79$ are the control
parameter values of the drive and response systems, respectively,
$\sigma$ characterizes the coupling strength between
systems~\cite{Pyragas:1996_WeakAndStrongSynchro}. Despite of the
fact that logistic map is the one-dimensional discrete system, it is
the etalon object of nonlinear dynamics demonstrating the wide
spectrum of interesting effects, and, therefore, it is typically
used to study different phenomena including chaotic synchronization. \alkor{Additionally, the logistic map belongs to the non-invertible discrete systems, for which the existence of the functional relation is not proven.}
Due to the one-dimensional character of interacting
systems~(\ref{eq:LogMaps}) the vectors in discussion given above
should be replaced by scalars, whereas all theoretical and
analytical findings remain correct.

\begin{figure}
\centerline{\includegraphics[scale=0.55]{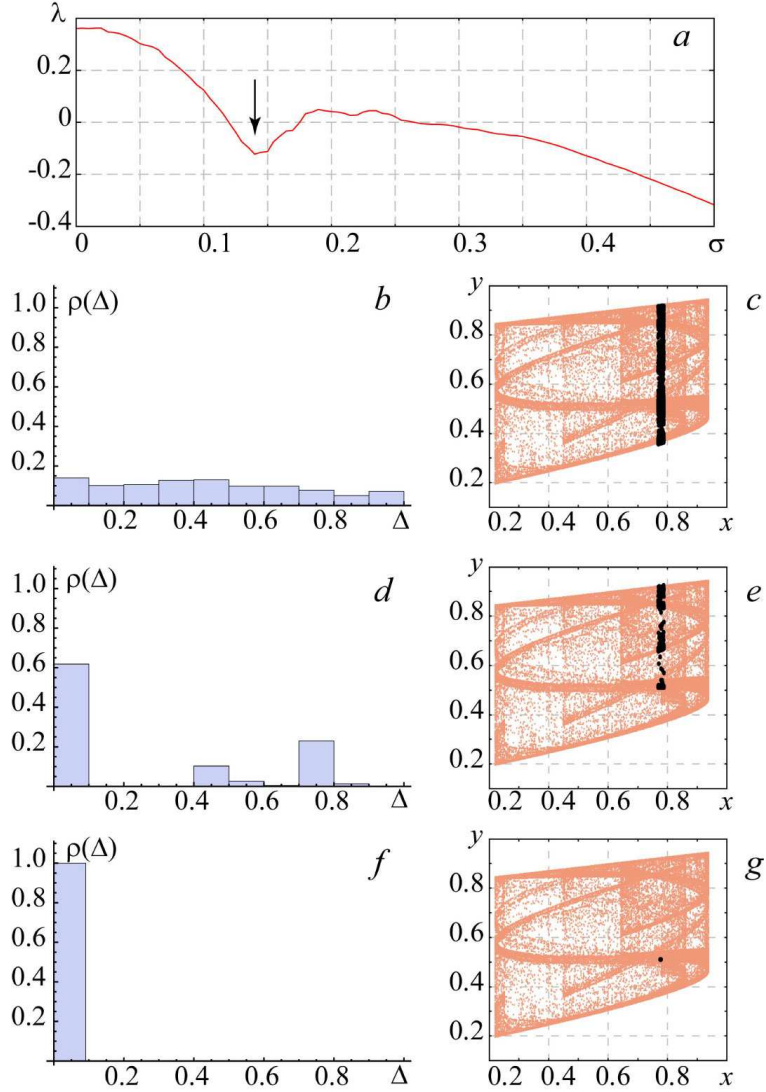}}
\caption{Dependence of the conditional Lyapunov exponent on the coupling parameter $\sigma$ (\textit{a}); histograms of the normalized differences $\Delta_{Ji}$ \alkor{built by $N=1000$ neighbor points} (\textit{b,d,f}) and $(x,y)$-planes (\textit{c,e,g})
for two unidirectionally coupled logistic maps (\ref{eq:LogMaps}) being in the generalized synchronization regime ($\sigma=0.14$, marked by arrow in this figure (\textit{a})) for the different lengths of the prehistory: $K=0$ \alkor{(the number of iteration used to achieve reasonable statistics is $L\sim3.3\times 10^4$)} (\textit{b,c}), $K=10$ \alkor{($L\sim5.5\times 10^6$)} (\textit{d,e}), $K=28$ \alkor{($L\sim 6.06\times 10^{10}$)} (\textit{f,g}). In Fig.~\ref{fgr:LogMaps},\textit{c,e,g} the points of the interacting systems satisfying requirement~(\ref{eq:InfinitesimalityAssumption}) are also shown}
\label{fgr:LogMaps}
\end{figure}

To detect the generalized synchronization regime we have computed
conditional Lyapunov exponent for system (\ref{eq:LogMaps}) with
further refinement of the threshold values with the help of the
auxiliary system method~\cite{Rulkov:1996_AuxiliarySystem}. In
Fig.~\ref{fgr:LogMaps} the dependence of the conditional Lyapunov
exponent on the coupling parameter $\sigma$ is shown. It is clearly
seen that conditional Lyapunov exponent is negative for
$\sigma\in[0.12;0.18]$ and $\sigma\geq 0.265$ that is the evidence
of the presence of the generalized synchronization regime in these
regions\footnote{This finding has been also verified with the help
of the auxiliary system approach.}. At that, the generalized
synchronization is close to the complete (strong) one if the
coupling parameter is a great enough, i.e. $\sigma\geq 0.265$,
whereas for $\sigma\in[0.12;0.18]$ the detected regime corresponds
to the so-called weak synchronization. It is clear that as in the
case of the flow systems in the strong generalized synchronization
there is no need to take prehistory into account because the drive
and response system states are related with each other by the simple
functional relation $y_n\approx
x_n$~\cite{Pyragas:1996_WeakAndStrongSynchro}. At the same time, the
case of weak synchronization (when $\sigma\in[0.12;0.18]$) demands
the additional investigation.

Without the loss of generality we fix the coupling parameter to be
${\sigma=0.14}$ that corresponds to the minimum negative value of the
conditional Lyapunov exponent (marked by arrow in
Fig.~\ref{fgr:LogMaps},\textit{a}). 
Having assumed the value of the accuracy in
Eq.~(\ref{eq:InfinitesimalityAssumption}) $\varepsilon=0.01$ we have
analyzed the influence of the length $K$ of the prehistory interval
on the points $\delta y_{Ji}$ and normalized
differences~(\ref{eq:NormDif}), with the reference point $x_N$ being
selected randomly\alkor{\footnote{It should be noted that the
quantitative value of the accuracy should be a small enough in
comparison with the amplitude of the signal from the system under
study and, at the same time, it should be a sufficiently large to
provide the reasonable statistics for a given time of
calculation.}}.
Obviously, when Eq.~(\ref{eq:VecorRelationship}) (or
Eq.~(\ref{eq:VecorFuncRelations})) is satisfied the  distribution of
normalized differences $\Delta_{Ji}$ should be the
$\delta$-function.

Fig.~\ref{fgr:LogMaps},\textit{b,d,f} illustrates the histograms of
the normalized differences $\Delta_{Ji}$ with the increase of the
length $K$ of the prehistory. \alkor{Histograms have been built by $N=1000$ neighbor points being closed to each other during all prehistory interval with the length $K$. To achieve such reasonable statistics for a given value of accuracy $\varepsilon$ we have done $L$ iterations the quantitative values of which are indicated in the caption to Fig.~\ref{fgr:LogMaps}.} In
Fig.~\ref{fgr:LogMaps},\textit{c,e,g} the $(x,y)$-planes
characterizing the drive and response system states for the selected
values of the control parameters are also shown. In each figure the
points $(x_{Ji},y_{Ji})$ of the interacting systems for which
requirement~(\ref{eq:InfinitesimalityAssumption}) is fulfilled are
also indicated. Fig.~\ref{fgr:LogMaps},\textit{b,c} corresponds to
the case when the prehistory is not taken into account at all, i.e.,
$K=0$. This consideration (without the prehistory) corresponds to
the traditional concept of the generalized synchronization generally
accepted hitherto. It is clearly seen that in this case the
normalized differences $\Delta_{Ji}$ are distributed uniformly over
the range $[0;1]$ (Fig.~\ref{fgr:LogMaps},\textit{b}), at that all
points in the phase space of the response system are also allocated
randomly in the wide range of the $y$-value variation
(Fig.~\ref{fgr:LogMaps},\textit{c}). So, we have to conclude that
Eq.~(\ref{eq:VecorFuncRelations}) fails and, as a consequence, the
traditional viewpoint on the generalized synchronization regime in
the discrete systems needs refining.

When the length of the prehistory increases, e.g., for $K=10$
(Fig.~\ref{fgr:LogMaps},\textit{d,e}), the separate peaks in the
normalized difference distribution are revealed (they are exist due
to the inhomogeneity of the chaotic attractor), although the points
$y_{Ji}$ in the phase space of the response system remain
distributed in the wide range of the $y$-value as before. Finally,
Fig.~\ref{fgr:LogMaps},\textit{f,g} illustrates the analogous
distributions for the optimal length of the prehistory interval
($K=28$). In this case the distribution of the normalized
differences $\Delta_{Ji}$ is the $\delta$-function
(Fig.~\ref{fgr:LogMaps},\textit{f}), and all considered system
states $(x_{Ji},y_{Ji})$ satisfying
requirement~(\ref{eq:InfinitesimalityAssumption}) are compressed
into small neighbourhood of the reference point $(x_{N},y_{N})$
(Fig.~\ref{fgr:LogMaps},\textit{g}).

So, having implemented the strategy developed above we can conclude
that \alkor{the relationship between states of the interacting logistic maps
involves the prehistory of the drive system
evolution in the same way as it has been revealed recently for the
systems with continuous time~\cite{Koronovskii:GSTubePRE2011}.}

\section{H\'enon maps}
\label{sct:Henon} As the second example we consider two mutually
coupled H\'enon maps. \alkor{The mutual type of coupling between interacting systems has been selected for the purpose of universality, i.e. the mutual coupling is so typical as the unidirectional one but analysis of the generalized synchronization regime is performed predominantly in the unidirectionally coupled dynamical systems.
There are also attempts to extend the concept of such phenomenon to the systems with a bidirectional type. For example, in our previous works~\cite{Moskalenko:GS_bidir_NDES2010,Moskalenko:GSbidirPRE2011} we have shown that the generalized synchronization regime in mutually coupled chaotic systems could be detected by the moment of transition of the second positive Lyapunov exponent in the field of the negative values.}

The system under study is given by:
\begin{equation}
\begin{array}{l}
x^1_{n+1}=f(x^1_n,x^2_n,a_x)+\sigma(f(y^1_n,y^2_n,a_y)-f(x^1_n,x^2_n,a_x)),\\
x^2_{n+1}=bx^1_n,\\
y^1_{n+1}=f(y^1_n,y^2_n,a_y)+\sigma(f(x^1_n,x^2_n,a_x)-f(y^1_n,y^2_n,a_y)),\\
y^2_{n+1}=by^1_n,\\
\end{array}
\label{eq:Henon}
\end{equation}
where $\mathbf{x}=(x^1,x^2)$ [$\mathbf{y}=(y^1,y^2)$] are the
vector-states of the first [second] system,
$f(x_1,x_2,a)=ax_1(1-x_1)+x_2$ is the  nonlinear function,
$a_x=3.16779$, $a_y=2.9$, $b=0.3$ are control parameters, $\sigma$
is the coupling
parameter~\cite{Pyragas:1997_CLEsFromTimeSeries,Pyragas:1998_GS}.
For the selected values of the control parameters generalized
synchronization determined by the moment of the transition of the
second positive Lyapunov exponent in the field of the negative
values~\cite{Moskalenko:GS_bidir_NDES2010,alkor:PierceBidir2011TPL} arises at
$\sigma\approx 0.035$.

\begin{figure}
\centerline{\includegraphics[scale=0.65]{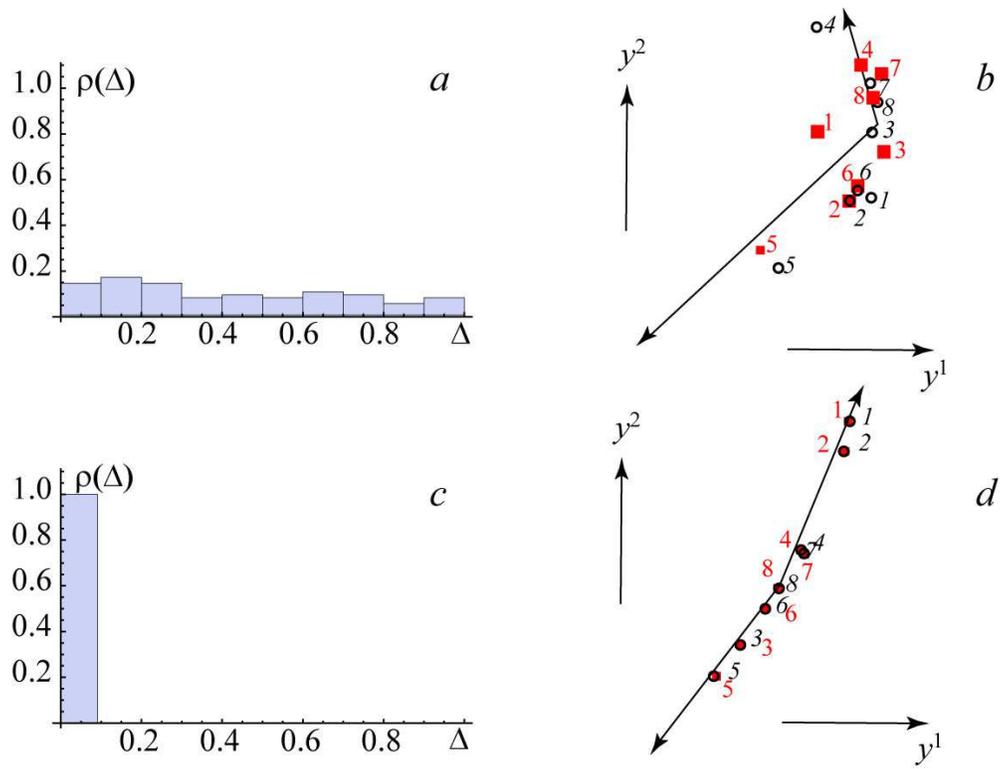}}
\caption{Histograms of the normalized differences $\Delta_{Ji}$ \alkor{built by $N=100$ neighbor points} (\textit{a,c}) and the vectors $\mathbf{y}_{Ji}$ (\textcolor{black}{$\circ$}) and $\mathbf{z}_{Ji}$ (\textcolor{red}{$\blacksquare$}) (\textit{b,d}) of the second H\'ennon map (\ref{eq:Henon}), $\sigma=0.2$, for the different lengths of the prehistory interval: $K=0$ \alkor{(the number of iterations used to achieve reasonable statistics is $L\sim5.9\times 10^4$)} (\textit{a,b}), $K=40$ \alkor{($L\sim6.2\times 10^{11}$)} (\textit{c,d}).
The numbers $i$ of the vectors  $\mathbf{y}_{Ji}$ and $\mathbf{z}_{Ji}$ are shown by the italic and regular fonts, respectively}
\label{fgr:Hennon}
\end{figure}

Now, we fix the coupling parameter to be $\sigma=0.2$ and apply the
strategy of the investigation of the generalized synchronization
essence (see above) to the system under study. For the chosen value
of the coupling parameter the weak generalized synchronization is
observed in system (\ref{eq:Henon}). As in the case of the logistic
maps we characterize the degree of closeness of the vectors
$\mathbf{y}_{Ji}$ and $\mathbf{z}_{Ji}$ by histograms of  the
normalized differences~(\ref{eq:NormDif}) \alkor{built by $N=100$ neighbor points. The quantitative values of iterations $L$ used to achieve such statistics are also shown in the caption of Fig.~\ref{fgr:Hennon}.} At the same time,
contrary to the case of the logistic maps considered above, the
system under study allows to visualize the behavior of the vectors
$\mathbf{y}_{Ji}$ and $\mathbf{z}_{Ji}$ in a plane. Therefore, in
Fig.~\ref{fgr:Hennon} along with the distributions of the normalized
differences $\Delta_{Ji}$ (Fig.~\ref{fgr:Hennon},\textit{a,c}) the
vectors $\mathbf{y}_{Ji}$ ($\circ$) and $\mathbf{z}_{Ji}$
(\textcolor{red}{$\blacksquare$})
(Fig.~\ref{fgr:Hennon},\textit{b,d}) of the second H\'enon map
(\ref{eq:Henon}) are shown. Fig.~\ref{fgr:Hennon},\textit{a,b}
corresponds to the case when all the nearest vectors
$\mathbf{x}_{Ji}$ satisfying
requirement~(\ref{eq:OneInfinitesimalityAssumption}) with
$\varepsilon=0.01$ are used (i.e., the length of the prehistory
interval is $K=0$ and Eq.~(\ref{eq:VecorFuncRelations}) is
examined), whereas Fig.~\ref{fgr:Hennon},\textit{b,c} refers to the
case when the prehistory of the length $K=40$ is taken into account
(in this case the requirement~(\ref{eq:InfinitesimalityAssumption})
with $\varepsilon=0.01$ is fulfilled and
Eq.~(\ref{eq:VecorRelationship}) is verified). It is clearly seen
that in the first case the normalized differences $\Delta_{Ji}$ are
distributed uniformly over the unit interval (as in the case of the
logistic maps considered above) and the vectors $\mathbf{z}_{Ji}$
and $\mathbf{y}_{Ji}$ differ from each other sufficiently that
testifies the failure of the presence of the functional relation
between the interacting system states. But, conversely, for the
second case when the prehistory is taken  into account the
distribution of $\Delta_{Ji}$ is a $\delta$-function and the
calculated vectors $\mathbf{z}_{Ji}$ are in the excellent agreement
with the vectors $\mathbf{y}_{Ji}$ of the second map that confirms
the theoretical predictions and results obtained above for the
unidirectionally coupled logistic maps. So, for in the two-dimensional
maps coupled mutually the vector states of the interacting chaotic
systems are not also related with each other by the \alkor{continuously differentiable} functional
relation and, again, the prehistory should be be taken into account.

\section{Weak and strong generalized synchronization}
\label{sct:WS&SSofGS}

In the final part  of our Letter we discuss briefly the existing
concept of the weak and strong synchronization (see,
e.g.,~\cite{Pyragas:1996_WeakAndStrongSynchro}) concerning the
generalized synchronization regime. As it has been mentioned above,
the strong and weak types of the generalized synchronization are
typically distinguished, according to the properties of the
functional relation between states of the systems. The onset of
generalized synchronization is believed to be characterized by an
unsmooth map $\mathbf{F}$ that becomes smooth only at sufficiently
large coupling strength. The synchronization types characterized by
a smooth and an unsmooth map were called a strong and weak
synchronization, respectively, with the complete and lag
synchronization being a particular case of strong synchronization.
This statement~\cite{Pyragas:1996_WeakAndStrongSynchro} was based on
the calculation of correlation dimension (and other characteristics)
of attractors in the phase space $\mathbb{D}\oplus\mathbb{R}$ (where
$\mathbb{D}$ and $\mathbb{R}$ are the phase spaces of the drive and
response oscillators, respectively)\footnote{For unsmooth map
$\mathbf{F}$ the dimension of a strange attractor in the whole phase
space $D\oplus R$ is supposed to be larger than the dimension of
driving attractor in $D$ space, whereas for smooth $\mathbf{F}$
these two dimensions must be equal.}.

Indeed, if one consider the attractor of two coupled logistic maps
in the $\mathbb{D}\oplus\mathbb{R}$-space (see
Fig.~\ref{fgr:LogMaps},\textit{c}), the fractal properties of it may
be easily revealed. At the same time, the fractality of the
relationship $\mathbf{F}$ between states of the interacting systems
is caused by the assumption of the
existence of the simple function relation~(\ref{eq:FunctRelGeneral})
between system states and neglecting the prehistory. As it has been
\alkor{discussed} above, the states of the interacting systems \alkor{may be not} related
with each other by the functional relation and the prehistory must
be taken into account. To introduce the prehistory into the
consideration in the $\mathbb{D}\oplus\mathbb{R}$-space only vectors
$\mathbf{y}_{Ji}$ must be used which satisfy
requirement~(\ref{eq:InfinitesimalityAssumption}) (see
Fig.~\ref{fgr:LogMaps},\textit{g}). As one can see, in this case all
considered system states $(x_{Ji},y_{Ji})$ satisfying
requirement~(\ref{eq:InfinitesimalityAssumption}) are compressed
into small neighbourhood of the reference point $(x_{N},y_{N})$, all
fractal properties disappear and the relation $\mathbf{F}$ between
the drive and response system states are smooth. The same conclusion
can be drawn not only for the logistic maps~(\ref{eq:LogMaps}) but
for the general case~(\ref{eq:EvolutionOperators}).

Nevertheless, the concept of the weak and strong types of the
generalized synchronization may be used in the improved form. This
improvement consists in the following. When the state of the second
system $\mathbf{y}_n$ depends on the prehistory (see
Eq.~(\ref{eq:FunctionalMaps})) with the length $K$ this type of the
synchronous dynamics should be considered as the weak generalized
synchronization. With the growth of the coupling strength the
required length $K$ of the prehistory decreases, and, for the certain
value of the coupling parameter $\sigma$ the length $K$ becomes
equal to zero and the complete synchronization regime is observed in
the system.
\alkor{Since for the unidirectionally coupled oscillators the length of the prehistory $K$ depends on the value of the maximum conditional Lyapunov exponent $\lambda_1^r$, the behavior of the prehistory length agrees well with the finding that the strong generalized synchronization occurs, when the maximum conditional Lyapunov exponent drops
below the minimum exponent of the drive system~\cite{Hunt:1997_DifferentiableGS_PRE}. When the maximum conditional Lyapunov exponent becomes less then the minimum exponent of the drive oscillator, the response system starts to be in some sense stiff enough to follow the external signal, whereas the required prehistory length $K$ is equal to zero.} In this case the states of the interacting systems are
related with each other by the \alkor{continuously differentiable} functional
relation~(\ref{eq:FunctRelMaps}) that should be considered as the
strong generalized synchronization.

So, the division of the generalized synchronization in discrete maps
on the  weak and strong ones is certainly justified. At the same
time, the difference between them is not determined by the type of
the relation $\mathbf{F}$ established between the interacting system
states (whether it is smooth or fractal), it is smooth in both
cases, at that in the case of strong synchronization the interacting
system states are related with each other by the functional
relation~(\ref{eq:FunctRelMaps}), whereas in for the weak one the
prehistory should be taken into account.

\section{Conclusions}
\label{sct:Conclusions} In conclusion, we have reported that as in
the case of the flow systems the concept of generalized
synchronization in discrete maps (coupled both unidirectionally and
mutually) needs refining, since \alkor{for the state vectors of the interacting
chaotic systems, in general, the prehistory should be taken into
account}. We have proposed the modification of the phase tube
approach applicable to the discrete maps and analyzed the essence of
the generalized synchronization by means of such technique. Obtained
results show that the division of the generalized synchronization
into the weak and the strong ones also needs refinement, i.e. in the
strong synchronization interacting system states are related with
each other by the \alkor{continuously differentiable} functional relation whereas in the weak one the
prehistory should be taken into account for the analysis of the
generalized synchronization regime. At that, both in the case of the
strong and weak synchronization relation established between the
interacting system states is smooth, and the so called
``fractality'' disappears when the appropriate consideration of the
prehistory is made.

At the same time, the found refinement of the generalized
synchronization in discrete maps  does not discard the majority of
the obtained hitherto results concerning its investigation. In
particular, the method of Lyapunov exponent computation and
auxiliary system approach remains valid as before as well as the
revealed mechanisms of the synchronous regime
arising~\cite{Aeh:2005_GS:ModifiedSystem,Harmov:2005_GSOnset_EPL}.
However, this refinement has an important fundamental significance
from the point of view of the understanding of the core mechanisms
of the considered phenomena and is supposed to give a strong
potential for new approaches and applications dealing with the
nonlinear systems. \alkor{Additionally, we expect that the phase tube approach gives a powerful detection and classification tool for the chaotic synchronization phenomenon study.}

\section{Acknowledgement}
\label{sct:Acknowledgement}

\alkor{We thank the Referees of our manuscript for useful comments
and remarks.} This work has been supported by Federal
special-purpose  programme ``Scientific and educational personnel of
innovation Russia'', Russian Foundation for Basic Research (project
12-02-00221), President's program (MK-672.2012.2) and  ``Dynasty''
Foundation.


\end{document}